\documentclass[a4paper]{article}
\usepackage{amsmath,amssymb,amsthm}
\usepackage{bbm}            
\usepackage{bm}             
\usepackage{graphicx,color}
\usepackage{mma}            
\usepackage{url}            

\newcommand{\rmd}{\,\mathrm{d}} 
\newcommand{\pa}{\partial}
\newcommand{\Ann}{\operatorname{Ann}}

\newcommand{\SUM}{\operatorname{SUM}}

\newcommand{\cmd}[1]{\textbf{#1}}

\newcommand{\Eq}{\mathcal{E}_q}

\renewcommand{\leq}{\leqslant}
\renewcommand{\geq}{\geqslant}

\newcommand{\Q}{\mathbbm Q}

\newcommand{\K}{\mathbbm K}
\newcommand{\F}{\mathbbm F}

\newcommand{\OO}{\mathbbm O}

\newcommand{\mmi}{\mathbbm i}                                          
\newcommand{\blank}{\underline{\;\;}}                                  
\newcommand{\pow}{\mathbin{\raisebox{-2.5pt}{\hbox{\large$\hat{}$}}}}  
\newcommand{\mmaref}[2]{{\sffamily #1[#2]}} 

\newtheorem{theorem}{Theorem}

\newtheorem{conjecture}[theorem]{Conjecture}
\newtheorem{definition}[theorem]{Definition}
\theoremstyle{definition}
\newtheorem{example}[theorem]{Example}

\newcommand{\thm}[1]{Theorem~\ref{#1}}

\renewcommand{\labelenumi}{(\arabic{enumi})}

\begin{document}

\title{Holonomic Tools for Basic Hypergeometric Functions}

\author{Christoph~Koutschan\thanks{Supported by the Austrian Science Fund (FWF): DK W1214.}\\
        Johann Radon Institute for Computational and\\
        Applied Mathematics (RICAM), Austrian Academy of Sciences\\
        Altenberger Stra\ss e 69, A-4040 Linz, Austria
\and
        Peter~Paule\thanks{Supported by the Austrian Science Fund (FWF): SFB F50-06.}\\
        Research Institute for Symbolic Computation (RISC)\\
        Johannes Kepler University\\
        Altenberger Stra\ss e 69, A-4040 Linz, Austria}

\maketitle

\hfill \emph{Dedicated to Professor Mourad Ismail}\\
\null\hfill \emph{at the occasion of his 70th birthday}
\bigskip

\begin{abstract}
With the exception of $q$-hypergeometric summation,
the use of computer algebra packages implementing Zeilberger's
``holonomic systems approach'' in a broader mathematical sense
is less common in the field of $q$-series and basic hypergeometric
functions. A major objective of this article is to popularize the
usage of such tools also in these domains. Concrete case studies
showing software in action introduce to the basic techniques.
An application highlight is a new computer-assisted proof
of the celebrated Ismail-Zhang formula, an important
$q$-analog of a classical expansion formula
of plane waves in terms of Gegenbauer polynomials.
\end{abstract}

\section{Introduction}
\label{sec.intro}

Quoting Knuth~\cite[p. 62]{KnuthDaylight14} the \emph{Concrete
  Tetrahedron}~\cite{KauersPaule11} is ``sort of the sequel to \emph{Concrete
  Mathematics}~\cite{GrahamKnuthPatashnik89}.''  Indeed, presenting
algorithmic ideas in connection with the symbolic treatment of combinatorical
sums, recurrences, and generating functions, it can be viewed as an
algorithmic supplement to~\cite{GrahamKnuthPatashnik89} directed at an
audience using computer algebra. Most of the methods under consideration fit
into the ``holonomic systems approach to special functions identities''
notably pioneered by Zeilberger~\cite{Zeilberger90}.

The authors of this article feel that in contrast to applications in the
domain of classical hypergeometric functions, the use of such methods and
tools is less common in the field of $q$-series and basic hypergeometric
functions. A major objective of this article is to popularize the holonomic
systems approach also in these domains. In order to illustrate some of the
basic (pun intended!) techniques, concrete case studies of software in action
are given.  To this end, computer algebra packages written in Mathematica are
used. These packages are freely downloadable (upon password request) by
following the instructions at
\centerline{\url{http://www.risc.uni-linz.ac.at/research/combinat/software}}

The article is structured as follows: 
In Section~\ref{sec.basic} the objects of a computational case study
are $q$-versions of modified Lommel
polynomials introduced by Ismail in~\cite{Ismail82}. To derive properties
of this polynomial family we apply $q$-holonomic computer
algebra tools for guessing, generalized telescoping, and
the execution of closure properties.

Section~\ref{sec.ann} introduces to the algebraic
language and concepts needed for an algorithmic
treatment of functions defined by mixed
($q$-)difference-differential equations. Following
Zeilberger's holonomic systems approach, special
functions are described by (generators of)
annihilating ideals in operator algebras.
Special function operations like addition, multiplication, integration,
or summation are lifted to operations on (the generators of)
these ideals. The HolonomicFunctions package implements this
algebraic/algorithmic framework. Gegenbauer polynomials
are used to show some basic features of the software.

Employing the algorithmic machinery described,
Section~\ref{sec.IZ} presents a new computer-assisted proof
of the celebrated Ismail-Zhang formula from~\cite{IsmailZhang94}. This important
identity is a $q$-analog of a classical expansion formula
of plane waves in terms of Gegenbauer polynomials and
involving Bessel functions. Ingredients of the $q$-analog
are the basic exponential function $\mathcal{E}_q(x; i \omega)$,
also introduced by Ismail and Zhang in~\cite{IsmailZhang94}, as well as
Jackson's second $q$-analog of the Bessel functions
$J_\nu(x)$ and $q$-Gegenbauer polynomials.

We want to mention explicitly that the task of computing an annihilating
operator for the series side of the Ismail-Zhang formula is leading to the
frontiers of what is computationally feasible today. To compute the operator
\cmd{annSumRHS} in \mmaref{In}{\ref{annSumRHS}}, we had to use recent
algorithmic developments~\cite{Koutschan10c} as well as human inspection and
trial and error to determine suitable denominators in a decisive preprocessing
step.

\section{Basic Bessel Functions and $q$-Lommel Polynomials}
\label{sec.basic}

We begin with basic Bessel functions considered by Ismail in~\cite{Ismail82}:
\[
  J_{\nu}^{(1)}(x;q) = \frac{(q^{\nu+1};q)_\infty}{(q;q)_\infty}
  \sum_{n=0}^\infty \frac{(-1)^n (x/2)^{\nu+2n}}{(q;q)_n (q^{\nu+1};q)_n},
  \qquad 0<q<1,
\]
where
\[
  (a;q)_0 = 1,\quad
  (a;q)_n = \prod_{j=0}^{n-1} (1-aq^j),\quad
  (a;q)_\infty = \prod_{j=0}^\infty (1-aq^j).  
\]

After opening a Mathematica session we load Riese's
package~\cite{PauleRiese97} which implements a $q$-version of
Zeilberger's ``fast'' Algorithm~\cite{Zeilberger90a}:
\begin{mma}
 \In <\!< \ |RISC\text{\'{}}qZeil\text{\'{}}|\\
\end{mma}
\newlength{\mmaboxwidth}
\setlength{\mmaboxwidth}{\textwidth}
\addtolength{\mmaboxwidth}{-32pt}
\setlength{\fboxsep}{3pt}
\setlength{\fboxrule}{1pt}
\vskip 5pt\noindent\hskip 24pt\fbox{\parbox{\mmaboxwidth}{\small
Package q-Zeilberger version 4.50 written by Axel Riese\\
Copyright 1992-2009, Research Institute for Symbolic Computation (RISC),\\
Johannes Kepler University, Linz, Austria}}
\vskip 8pt
\noindent
The package provides the $q$-rising factorials via the \cmd{qPochhammer} command, i.e.,
\cmd{qPochhammer}$[a,q,k]:=(a;q)_k$ and \cmd{qPochhammer}$[a,q]:=(a;q)_\infty$.
For better readability we set
\begin{mma}
  \In (a\blank)_{k\blank}:=|qPochhammer|[a,q,k]\\
\end{mma}
\noindent
A recurrence for $J_\nu^{(1)}(x;q)$ ($=:\SUM[\nu]$) is computed as follows:
\begin{mma}
  \In |qZeil|\left[\frac{\bigl(q^{\nu+1}\bigr)_{\infty}(-1)^n(x/2)^{\nu+2n}}{(q)_{\infty}(q)_n\bigl(q^{\nu+1}\bigr)_n},
    \{n,0,\infty\},\nu,2\right]\rule[-20pt]{0pt}{0pt}\\
  \Warning{qZeil::natbounds} Assuming appropriate convergence.\\
  \Out \text{SUM}[\nu]=q^{1-\nu} \bigl(-\text{SUM}[\nu-2]\bigr)-\frac{2 \left(1-q^{1-\nu}\right) \text{SUM}[\nu-1]}{x}\\
  \label{qZeil}
\end{mma}
\noindent
This corresponds exactly to (1.18), $k=1$, in~\cite{Ismail82}.
Setting $r_1^{(\nu)}(x):=2(1-q^\nu)x$ we rewrite the previous output \mmaref{Out}{\ref{qZeil}} as
\begin{equation}\label{eq.1}
  q^\nu \SUM[\nu+1] = r_1^{(\nu)}\Bigl(\frac1x\Bigr)\SUM[\nu] - \SUM[\nu-1].
\end{equation}
By iterating this recurrence, one produces a sequence $r_1^{(\nu)}(x),r_2^{(\nu)}(x),r_3^{(\nu)}(x),\dots$
of polynomials such that
\begin{align}
  \label{eq.2}
  q^\nu q^{\nu+1} \SUM[\nu+2] &= r_2^{(\nu)}\Bigl(\frac1x\Bigr) \SUM[\nu] - r_1^{(\nu+1)} \Bigl(\frac1x\Bigr) \SUM[\nu-1],\\
  \label{eq.3}
  q^\nu q^{\nu+1} q^{\nu+2} \SUM[\nu+3] &= r_3^{(\nu)} \Bigl(\frac1x\Bigr) \SUM[\nu] - r_2^{(\nu+1)} \Bigl(\frac1x\Bigr) \SUM[\nu-1],
\end{align}
and so on.
In other words, setting $r_0^{(\nu)}(x):=1$, the polynomial sequence
$\bigl(r_n^{(\nu)}(x)\bigr){}_{n\geq0}$ determined this way satisfies the
relation
\begin{equation}
  q^{n\nu+n(n-1)/2} J_{\nu+n}^{(1)}(x;q) =
  r_n^{(\nu)}\Bigl(\frac1x\Bigr) J_\nu^{(1)}(x;q) - r_{n-1}^{(\nu+1)}\Bigl(\frac1x\Bigr) J_{\nu-1}^{(1)}(x;q),
  \quad n\geq1.
\end{equation}

This is recurrence (1.19) for $k=1$ in~\cite{Ismail82}. As noted ibid. the
polynomials $r_n^{(\nu)}(x)$ are $q$-versions of the modified Lommel polynomials.
The goal of the present case study is to illustrate how computer algebra tools
can be used to find out more about the polynomials $r_n^{(\nu)}(x)$.

\subsection{Guessing a $q$-holonomic recurrence}

First, by iterating recurrence~\eqref{eq.1} as in~\eqref{eq.2}
and~\eqref{eq.3}, we compute the seven initial polynomials
$1,r_1^{(\nu)}(x),r_2^{(\nu)}(x),\ldots,r_6^{(\nu)}(x)$ and store them in a list
(not shown in full detail here for space reasons):
\begin{mma}
 \In |rpolys|=\{1,2 (1-q^\nu) x,(4-4 q^\nu-4 q^{1+\nu}+4 q^{1+2 \nu})x^2-q^\nu  ,
  -2 (-1+q^{1+\nu}) x (-q^\nu-q^{1+\nu}+4 x^2-4 q^\nu x^2-4 q^{2+\nu} x^2+4 q^{2+2 \nu} x^2), \dots\};\\
\end{mma}
\noindent
As described in~\cite{KauersKoutschan09}, the package
\begin{mma}
 \In <\!< \ |RISC\text{\'{}}qGeneratingFunctions\text{\'{}}|\\
\end{mma}
\vskip 5pt\noindent\hskip 24pt\fbox{\parbox{\mmaboxwidth}{\small
qGeneratingFunctions Package version 1.9.1\\
written by Christoph Koutschan\\
Copyright 2006-2015, Research Institute for Symbolic Computation (RISC),\\
Johannes Kepler University, Linz, Austria}}
\vskip 8pt
\noindent
can be used to guess a recursive pattern for the
sequence~$\bigl(r_n^{(\nu)}(x)\bigr){}_{n\geq0}$.  To do so, we execute
\begin{mma}
  \In |srec|=|QREGuess|[|rpolys|,s[n]]\rule[-8pt]{0pt}{0pt}\\
  \Warning{QREGuess::data} Not enough data. The result might be wrong.\\
  \Out \bigl\{-2 q x \, s[n-1] \left(q-q^{n+\nu}\right)+s[n-2] q^{n+\nu}+q^2 s[n]=0,\linebreak
    \phantom{\bigl\{}s[0]=1,s[1]=2 x \left(1-q^\nu\right)\bigr\}\\
  \label{srec}
\end{mma}
\noindent
Ignoring the warning, and observing that when using as input more than $7$
polynomials the guessed recurrence remains stable, the output (i.e., the
automatic guess) can be interpreted as a conjecture (it corresponds to (1.20)
for $k=1$ in~\cite{Ismail82}).
\begin{conjecture}\label{conj.1}
Let $s_n^{(\nu)}(x),n\geq0$, be the sequence uniquely defined by the
recurrence in {\rm\mmaref{Out}{\ref{srec}}}. Then $r_n^{(\nu)}(x)=s_n^{(\nu)}(x)$
for all $n\geq0$.
\end{conjecture}
\begin{definition}
A sequence $(a_n)_{n\geq0}$ that satisfies a linear recurrence with coefficients
being polynomials in $q^n$ with coefficients in a field $\K(q)$ is called
$q$-holonomic.
\end{definition}
In (computational) applications the coefficient field $\K(q)$ is a rational
function field; usually $\K$ is a transcendental extension
$\K=\Q(a,b,c,\ldots)$ of~$\Q$ containing parameters $a$, $b$, $c$, and so
on. In our example, $\K=\Q(q^\nu)$.

As pointed out in \cite{Ismail82} Conjecture~\ref{conj.1} can be proved by
straightforward induction. In order to introduce various aspects of computer
algebra we present an algorithmic proof exploiting \emph{holonomic closure
  properties}.

\subsection{Proof of Conjecture~\ref{conj.1} and $q$-holonomic closure properties}

Iterating recurrence~\eqref{eq.1} as in~\eqref{eq.2} and~\eqref{eq.3} uniquely
determines the polynomials~$r_n^{(\nu)}$. Hence to prove Conjecture~\ref{conj.1}
it suffices to prove
\begin{equation}\label{eq.5}
  q^{n\nu+n(n-1)/2} J_{\nu+n}^{(1)}(x;q) =
  s_n^{(\nu)}\Bigl(\frac1x\Bigr) J_\nu^{(1)}(x;q) - s_{n-1}^{(\nu+1)}\Bigl(\frac1x\Bigr) J_{\nu-1}^{(1)}(x;q),
  \quad n\geq1,
\end{equation}
where $s_n^{(\nu)}(x)$ is the sequence uniquely defined by the recurrence
\mmaref{Out}{\ref{srec}} together with the initial values
$s_0^{(\nu)}:=1=r_0^{(\nu)}$ and $s_1^{(\nu)}(x):=2(1-q^\nu)x=r_1^{(\nu)}(x)$.

First we call the \cmd{qZeil} package to obtain a recurrence with respect to~$n$
for the left side of~\eqref{eq.5}:
\begin{mma}
  \In |recLHS|=|qZeil|\biggl[q^{n\nu+n((n-1)/2)} \frac{(q^{\nu+n+1})_{\infty}(-1)^j(x/2)^{\nu+n+2j}}{(q)_j(q^{\nu+n+1})_j},\linebreak
    \{j,0,\infty\},n,2,\{\nu\}\biggr]/.n\to n+2\rule[-16pt]{0pt}{0pt}\\
  \Warning{qZeil::natbounds} Assuming appropriate convergence.\\
  \Out \text{SUM}[n+2]=\frac{2 \text{SUM}[n+1] \left(1-q^{n+\nu+1}\right)}{x}-\text{SUM}[n] q^{n+\nu}\\
  \label{recLHS}
\end{mma}
In what follows it will be convenient to work directly with operators. To this end we load
\begin{mma}
 \In <\!< \ |RISC\text{\'{}}HolonomicFunctions\text{\'{}}|\\
\end{mma}
\vskip 5pt\noindent\hskip 24pt\fbox{\parbox{\mmaboxwidth}{\small
HolonomicFunctions Package version 1.7.1 (09-Oct-2013)\\
written by Christoph Koutschan\\
Copyright 2007-2013, Research Institute for Symbolic Computation (RISC),\\
Johannes Kepler University, Linz, Austria}}
\vskip 8pt The following procedure writes the recurrence \cmd{recLHS}, which
is \mmaref{Out}{\ref{recLHS}}, into an operator \cmd{opLHS} which
annihilates $q^{n\nu+n(n-1)/2}J_{n+\nu}^{(1)}(x;q)$, the left-hand side
of~\eqref{eq.5}:
\begin{mma}
  \In |opLHS|=|ToOrePolynomial|[|recLHS|,|SUM|[n],|OreAlgebra|[|QS|[|N|,q^n]]]\\
  \Out S_{\!N,q}^2+\Bigl(\frac{2 N q^{\nu+1}}{x}-\frac{2}{x}\Bigr) S_{\!N,q}+N q^\nu\\
  \label{opLHS}
\end{mma}
\noindent
The notation becomes clear by comparison to \cmd{recLHS}: $N$ stands for
$q^n$, and $S_{\!N,q}$ denotes the shift operator with respect to~$n$. For
instance, $S_{\!N,q}F(n)=F(n+1)$, $S_{\!N,q}N=qNS_{\!N,q}$, and
$S_{\!N,q}^2F(n)=F(n+2)$.

Calling the same procedure from the \cmd{HolonomicFunctions} package, we obtain
operator forms of the recurrences of the two parts on the right-hand side
of~\eqref{eq.5}:
\begin{mma}
  \In |srec1|=|srec|[[1]]/.\{x\to 1/x,n\to n+2\}\\
  \Out -\frac{2 q s[n+1] \left(q-q^{n+\nu+2}\right)}{x}+s[n] q^{n+\nu+2}+q^2 s[n+2]=0\\
\end{mma}
\noindent
This recurrence for the $s_n^{(\nu)}(1/x)$ then is rewritten as an annihilating
operator of $s_n^{(\nu)}\left(\frac1x\right)J_\nu^{(1)}(x;q)$:
\begin{mma}
  \In |op1RHS|=\{|ToOrePolynomial|[|srec1|,s[n],|OreAlgebra|[|QS|[|N|,q^n]]]\}\\
  \Out \biggl\{q^2 S_{\!N,q}^2+\Bigl(\frac{2 N q^{\nu+3}}{x}-\frac{2 q^2}{x}\Bigr) S_{\!N,q}+N q^{\nu+2}\biggr\}\\
\end{mma}
\noindent
Analogously we obtain an annihilating operator of $-s_{n-1}^{(\nu+1)}\left(\frac1x\right) J_\nu^{(1)}(x;q)$:
\begin{mma}
  \In |srec2|=|srec|[[1]]/.\{x\to 1/x,n\to n+2,\nu\to \nu+1\}\\
  \Out -\frac{2 q s[n+1] \left(q-q^{n+\nu+3}\right)}{x}+s[n] q^{n+\nu+3}+q^2 s[n+2]=0\\
  \In |op2RHS|=\{|ToOrePolynomial|[|srec2|,s[n],|OreAlgebra|[|QS|[|N|,q^n]]]\}\\
  \Out \biggl\{q^2 S_{\!N,q}^2+\Bigl(\frac{2 N q^{\nu+4}}{x}-\frac{2 q^2}{x}\Bigr) S_{\!N,q}+N q^{\nu+3}\biggr\}\\
\end{mma}
\noindent
By constructively utilizing the holonomic closure properties we compute an
operator annihilating $s_n^{(\nu)}\left(\frac1x\right)J_\nu^{(1)}(x;q)-s_{n-1}^{(\nu+1)}\left(\frac1x\right) J_\nu^{(1)}(x;q)$:
\begin{mma}
  \In |opRHS|=|DFinitePlus|[|op1RHS|,|op2RHS|][[1]] // |Factor|\\
  \Out x^2 S_{\!N,q}^4+2 (q+1) x \left(N q^{\nu+3}-1\right) S_{\!N,q}^3+q \bigl(4 N^2 q^{2 \nu+5}+N x^2 q^{\nu+1}+N x^2 q^{\nu+2}-{}\linebreak
    4 N q^{\nu+2}-4 N q^{\nu+3}+4\bigr) S_{\!N,q}^2+2 N (q+1) x q^{\nu+2} \left(N q^{\nu+2}-1\right) S_{\!N,q}+N^2 x^2 q^{2 \nu+3}\\
\end{mma}
\noindent
As often when applying holonomic closure properties this operator is not equal
to \cmd{opLHS}, but a left multiple of it:
\[
  \text{\cmd{opRHS}} = \left(x^2S_{\!N,q}^2 + 2q(Nq^{\nu+3}-1)xS_{\!N,q} + Nq^{\nu+3}x^2\right) \text{\cmd{opLHS}}.
\]
With the \cmd{HolonomicFunctions} package this factorization can be found as follows:
\begin{mma}
  \In |LMultiple|=|OreReduce|[|opRHS|,\{|opLHS|\},|Extended|\to |True|] \\ 
  \Out \bigl\{0,1,\bigl\{x^2 S_{\!N,q}^2+2 q x \left(N q^{\nu+3}-1\right) S_{\!N,q}+N x^2 q^{\nu+3}\bigr\}\bigr\}\\
  \In |OreTimes|[|LMultiple|[[3,1]],|opLHS|]\\
  \Out x^2 S_{\!N,q}^4+2 (q+1) x \left(N q^{\nu+3}-1\right) S_{\!N,q}^3+q \bigl(4 N^2 q^{2 \nu+5}+N x^2 q^{\nu+1}+N x^2 q^{\nu+2}-{}\linebreak
    4 N q^{\nu+2}-4 N q^{\nu+3}+4\bigr) S_{\!N,q}^2+2 N (q+1) x q^{\nu+2} \left(N q^{\nu+2}-1\right) S_{\!N,q}+N^2 x^2 q^{2 \nu+3}\\
\end{mma}
Summarizing, we have shown that both sides of~\eqref{eq.5} satisfy a recurrence of order~$4$ with respect to~$n$.
Hence the proof of Conjecture~\ref{conj.1} is completed by verifying~\eqref{eq.5} for $n=1,2,3,4$ which amounts to 
checking $r_n^{(\nu)}(x)=s_n^{(\nu)}(x)$ for $n=0,1,2,3$.

\subsection{$q$-holonomic functions and generalized telescoping}

In order to gain more insight into the structure of the polynomials
$s[n]:=s_n^{(\nu)}(x)$ we look at the generating function
\[
  F(t) := \sum_{n=0}^\infty s[n]t^n = \sum_{j=0}^\infty s_n^{(\nu)}(x) t^n.
\]
Taking as input the recurrence \cmd{srec}, which including the initial
values is a unique presentation of $\bigl(s_n^{(\nu)}(x)\bigr){}_{n\geq0}$, we
first derive a $q$-difference equation for $F(t)$ by calling a procedure from
the \cmd{qGeneratingFunctions} package:
\begin{mma}
  \In |qDiffEq|=|QRE2SE|[|srec|,s[n],|F|[t]]\\
  \Out \bigl\{t q^\nu (t+2 x) F[qt]+F[t] (1-2 t x)-1=0,\langle 1\rangle [F[t]]=1\bigr\}\\
  \label{qDiffEq}
\end{mma}
Here $\langle1\rangle[F[t]]=1$ stands for $F[0]=1$. In view of the $q$-shift
operator $S_{t,q}^kF(t)=F(q^kt)$, $q$-difference equations like this are also
called $q$-shift equations.
\begin{definition}
A function $F(t)$ that satisfies a linear $q$-difference equation with
coefficients being polynomials in $t$ with coefficients in a field $\K(q)$ is
called $q$-holonomic.
\end{definition}
As noted in \cite{Ismail82} \cmd{qDiffEq}, which is
\mmaref{Out}{\ref{qDiffEq}}, can be iterated to obtain an explicit
series representation for~$F(t)$:
\begin{align*}
  (1-2tx) F(t) &= 1 - 2xtq^\nu\Bigl(1+\frac12\frac tx\Bigr) F(qt) \\
  &= 1 - 2xtq^\nu\frac{1+\frac12\frac tx}{1-2txq} +
    q(2xtq^\nu)^2 \frac{1+\frac12\frac tx}{1-2txq} \frac{1+\frac12\frac tx q}{1-2txq^2} F(q^2t) + \ldots
\end{align*}
In the limit this iteration process results in \cite[(3.3)]{Ismail82}:
\begin{equation}\label{eq.6}
  F(t) = \sum_{j=0}^\infty \frac{(-2xtq^\nu)^j \left(-\frac12\frac tx\right){}_j}{\bigl(2xt\bigr){}_{j+1}} q^{j(j-1)/2}.
\end{equation}
The series presentation~\eqref{eq.6} can be used to derive a
$q$-hypergeometric sum representation for the~$s_n^{(\nu)}(x)$ using two
versions of the $q$-binomial theorem, the finite version by Gau\ss\ and the
infinite one by Heine; see (3.3) and (3.6) in~\cite{Ismail82}.

To illustrate further functionalities of the \cmd{HolonomicFunctions} package we
derive a $q$-difference equation for the right-hand side of~\eqref{eq.6}:
\begin{mma}
  \In |CreativeTelescoping|\biggl[\frac{(-2xtV)^j\bigl(-\frac12t/x\bigr){}_j}{(2xt)_{j+1}}q^{j(j-1)/2},\linebreak
    |QS|[J,q^j]-1,|QS|[t,q^T]\biggr]/.V\to q^\nu // |Factor|\\
  \Out \bigl\{\{-t q^\nu (t+2 x) S_{t,q}+(2 t x-1)\},\{2 t x-1\}\bigr\}\\
  \label{qct}
\end{mma}
\noindent
Setting
\[
  g_j(t) := \frac{(-2xtq^\nu)^j \left(-\frac12\frac tx\right){}_j}{\bigl(2xt\bigr){}_{j+1}} q^{j(j-1)/2}
\]
and $\Delta_jF(j):=F(j+1)-F(j)$, the output constitutes the solution of a
generalized telescoping problem and means that
\[
  q^\nu t (t+2x) g_j(qt) - (2tx-1)g_j(t) = \Delta_j(2tx-1)g_j(t).
\]
Summing this from $j=0$ to $j=\infty$, and setting the right-hand side
of~\eqref{eq.6} to $G(t)$, gives
\[
  q^\nu t (t+2x) G(qt) - (2tx-1)G(t) = (2tx-1)\left(g_\infty(t)-\frac{1}{1-2xt}\right).
\]
Noting that $g_\infty(t)=0$ (as a formal power series in~$q$, or
analytically taking $|q|<1$) we obtain \cmd{qDiffEq}; i.e., $G(t)$ satisfies
the same $q$-difference equation as for~$F(t)$.

As in the case $q=1$~\cite[Thm. 7.1]{KauersPaule11} there is a simple but
important connection between $q$-holonomic generating functions and their
coefficient sequences:
\begin{theorem}
\[
  F(t) = \sum_{n=0}^\infty a_nt^n \text{ is $q$-holonomic}
  \quad\iff\quad
  \bigl(a_n\bigr){}_{n\geq0} \text{ is $q$-holonomic}.
\]
\end{theorem}
One direction of the theorem has been exploited above when deriving
\cmd{qDiffEq} from \cmd{srec}. The inverse direction, this means, to compute
from the $q$-difference equation for $F(t)$ a $q$-recurrence for the coefficient
polynomials $s(n)=s_n^{(\nu)}(x)$, is done as follows:
\begin{mma}
  \In |QSE2RE|[|qDiffEq|,|F|[t],s[n]]\\
  \Out \left\{q^2 s[n]=2 q x\, s[n-1] \left(q-q^{n+\nu}\right)-s[n-2] q^{n+\nu},s[0]=1,s[1]=-2 x \left(q^\nu-1\right)\right\}\\
\end{mma}
\noindent
The output is \cmd{srec}, the $q$-recurrence \mmaref{Out}{\ref{srec}}
for the modified $q$-Lommel polynomials.

\section{Interlude: annihilating ideals of operators}
\label{sec.ann}

In order to state, in an algebraic language, the concepts that are introduced
in this section, and for writing mixed ($q$-)difference-differential equations in a
concise way, the following operator notation is employed: let $D_{\!x}$ denote
the partial derivative operator with respect to~$x$ ($x$ is then called a
\emph{continuous variable}), $S_n$ the forward shift operator with respect to~$n$
($n$ is then called a \emph{discrete variable}), and $S_{t,q}$ the $q$-shift operator
with respect to~$t$. More precisely, these operator symbols act on a function~$f$ by
\[
  D_{\!x}f = \frac{\pa f}{\pa x},
  \qquad
  S_nf = f\big|_{n\to n+1},
  \quad\text{and}\quad
  S_{t,q}f = f\big|_{t\to qt}.
\]
The operator notation allows us to translate linear homogeneous
($q$-)difference-differential equations into polynomials in the operator
symbols $D_{\!x}$, $S_n$, $S_{t,q}$, etc., with coefficients in some
field~$\F$. Typically, $\F$ is a rational function field in the
variables $x$, $n$, $t$, $q$, etc. For example, the equation
\[
  \frac{\pa}{\pa x}f(k,n+1,x,y) + n\frac{\pa}{\pa y}f(k,n,x,y) + xf(k+1,n,x,y) - f(k,n,x,y) = 0
\]
turns into $Pf(k,n,x,y)=0$, where $P$ is the operator $D_{\!x}S_n + nD_{\!y} +
xS_{\!k} - 1$.  An example in the $q$-case is the annihilating operator
\cmd{opLHS}, given in \mmaref{Out}{\ref{opLHS}}, for
$q^{n\nu+n(n-1)/2}J_{n+\nu}^{(1)}(x;q)$,
\begin{equation}\label{eq.3.1}
  J := S_{\!N,q}^2 + \left(-\frac2x+\frac{2Nq^{\nu+1}}{x}\right)S_{\!N,q} + Nq^\nu,
\end{equation}
which is a polynomial in the $q$-shift operator $S_{\!N,q}$ whose coefficients
are elements of the rational function field $\Q(q,q^\nu,N,x)$ where
$N=q^n$.  Note that in general the ring $\F[D_{\!x},S_n,S_{t,q},\dots]$ is
not commutative: coefficients from $\F$ do not commute with the ``variables''
$D_{\!x}$, $S_n$, $S_{t,q}$, etc. For instance, for some $a(x,n,t)\in \F$ one has
\begin{align*}
  D_{\!x}\cdot a(x,n,t) &= a(x,n,t)\cdot D_{\!x} + \frac{\pa}{\pa x} a(x,n,t), \\
  S_n\cdot a(x,n,t) &= a(x,n+1,t)\cdot S_n,\\
  S_{t,q}\cdot a(x,n,t) &= a(x,n,qt)\cdot S_{t,q}.
\end{align*}
Such non-commutative rings of operators are called \emph{Ore algebras}; more
precise definitions and properties of such algebras can be found in~\cite{Koutschan09}.

\begin{example}
We demonstrate how arithmetic operations in an Ore algebra can be used
to compute the polynomials $r_i^{(\nu)}$ for $i=1,2,3,\dots$. For this purpose
let us convert the recurrence \mmaref{Out}{\ref{qZeil}} into an operator:
\begin{mma}
  \In |op|=|Factor|[|ToOrePolynomial|[|SUM|[\nu]==-q^{1-\nu}\,|SUM|[\nu-2]-\linebreak
      2(1-q^{1-\nu}\,|SUM|[\nu-1])/x,|SUM|[\nu],|OreAlgebra|[|QS|[V,q^\nu]]]]\\
  \Out S_{V,q}^2+\frac{2 (q V-1)}{q V x} S_{V,q}+\frac{1}{q V}\\
\end{mma}
\noindent
Then iterating this recurrence according to~\eqref{eq.2} and~\eqref{eq.3}
corresponds to reducing the operator $V (qV)\cdots (q^{i-1}V) S_{V,q}^i$,
which encodes the left-hand side, with the previously defined \cmd{op};
the result is an operator that corresponds to the right-hand side.
Its leading coefficient is precisely the desired $r_i^{(\nu+1)}(1/x)$
(note that the symbol \cmd{**} stands for noncommutative multiplication).
\begin{mma}
  \In |Table|[|LeadingCoefficient|[|OreReduce|[(q^{i(i-1)/2}V^{i-1})\,|**|\,|QS|[V,q^\nu]^i,\linebreak
        \{|op|/.x\to 1/x\}]]/.V\to V/q,\{i,1,4\}]\\
  \Out \bigl\{1,-2 (V-1) x,4 q V^2 x^2-4 q V x^2-4 V x^2-V+4 x^2,\linebreak
    \phantom{\bigl\{}-2 x (q V-1) \bigl(4 q^2 V^2 x^2-4 q^2 V x^2-q V-4 V x^2-V+4 x^2\bigr)\bigr\}\\
\end{mma}
\end{example}

We define the \emph{annihilator} (with respect to some Ore algebra~$\OO$) of a
function~$f$ by:
\[
  \Ann_{\OO}(f) := \{P\in\OO\mid Pf=0\}.
\]
It can easily be seen that $\Ann_{\OO}(f)$ is a left ideal in~$\OO$. Every left
ideal~$I\subseteq\Ann_{\OO}(f)$ is called an \emph{annihilating ideal} for~$f$.
For example, the operator~$J$ given in~\eqref{eq.3.1} is an element of
$\Ann_{\OO}\bigl(q^{n\nu+n(n-1)/2}J_{n+\nu}^{(1)}(x;q)\bigr)$ with
$\OO=\F[S_{\!N,q}]=\Q(q,q^\nu,N,x)[S_{\!N,q}]$. Actually, it is the unique (up
to multiplication by elements from~$\F$) generator of that principal left ideal.

\begin{definition}
  Let $\OO=\F[\dots]$ be an Ore algebra.  A function~$f$ is called
  \emph{$\pa$-finite} with respect to~$\OO$ if $\OO/\Ann_{\OO}(f)$ is a finite-dimensional
  $\F$-vector space. The dimension of this vector space is called the
  \emph{(holonomic) rank} of~$f$ with respect to~$\OO$.
\end{definition}

\begin{example}\label{ex.Gegen1}
The following \cmd{HolonomicFunctions} procedure delivers the annihilator of the
Gegenbauer (also called ultraspherical) polynomials $C_m^{(\nu)}(x)$:
\begin{mma}
  \In |annG|=|Annihilator|[|GegenbauerC|[m,\nu,x],\{|S|[m],|S|[\nu],|Der|[x]\}]\\
  \Out \bigl\{2 \nu S_{\nu}-x D_{\!x}+(-m-2 \nu),\,(m+1) S_{m}+(1-x^2) D_{\!x}+(-m x-2 \nu x),\linebreak
    \phantom{\bigl\{}(x^2-1) D_{\!x}^2+(2 \nu x+x) D_{\!x}+(-m^2-2 m \nu)\bigr\}\\
  \label{annG}
\end{mma}
\noindent
This means, these three elements generate
$I:=\Ann_{\OO}\bigl(C_m^{(\nu)}(x)\bigr)$ as a left ideal in the operator algebra
$\OO=\F[S_m,S_\nu,D_x]$ with $\F=\Q(m,\nu,x)$.  Their leading monomials are
$S_\nu$, $S_m$, and $D_x^2$, which shows that the function $C_m^{(\nu)}(x)$ is
$\partial$-finite with respect to~$\OO$ and in particular:
\[
  \operatorname{rank}_{\OO}\bigl(C_m^{(\nu)}(x)\bigr) = \dim_{\F}\bigl(\OO / I\bigr) = 2.
\]
\end{example}

In the holonomic systems approach, the data structure for representing
functions is an annihilating ideal (given by a finite set of generators) plus
initial values. When working with (left) ideals, we make use of \emph{(left)
  Gr\"obner bases}~\cite{Buchberger65,KandrirodyWeispfenning90} which are an
important tool for executing certain operations (e.g., the ideal membership
test) in an algorithmic way.

For functions annilated by univariate operators from the Ore algebras
$\F[S_n]$ or $\F[D_x]$ or $\F[S_{\!N,q}]$, the notions of $\pa$-finite and
($q$-)holonomic coincide. Despite being closely related to being
$\pa$-finite, for functions annihilated by multivariate Ore operators the
definition of holonomic is much more technical. In general, the holonomic
property reflects certain elimination properties of annihlating operators
which are required for summation and integration of special functions. 

Without proof we state the following theorem about \emph{closure properties}
of $\pa$-finite functions; its proof can be found in
\cite[Chap. 2.3]{Koutschan09}.  We remark that all of them are algorithmically
executable, and the algorithms work with the above mentioned data structure.

\begin{theorem}\label{thm.clprop}
Let $\OO$ be an Ore algebra and let~$f$ and~$g$ be $\pa$-finite with respect to~$\OO$
of rank $r$ and $s$, respectively. Then
\renewcommand{\labelenumi}{(\roman{enumi})}
\setlength{\leftmargini}{2.5em}
\begin{enumerate}
\itemsep 0.5em
\item $f+g$ is $\pa$-finite of rank $\leq r+s$.
\item $f\cdot g$ is $\pa$-finite of rank $\leq rs$.
\item $f^2$ is $\pa$-finite of rank $\leq r(r+1)/2$.
\item $Pf$ is $\pa$-finite of rank $\leq r$ for any $P\in\OO$.
\item $f|_{x\to A(x,y,\dots)}$ is $\pa$-finite of rank $\leq rd$ if $x,y,\dots$
  are continuous variables and if the algebraic function $A$ satisfies a
  polynomial equation of degree~$d$.
\item $f|_{n\to A(n,k,\dots)}$ is $\pa$-finite of rank $\leq r$ if $A$ is an
  integer-linear expression in the discrete variables $n,k,\dots$.
\end{enumerate}
\end{theorem}

The bounds on the ranks are generically sharp. For example, the operator
\cmd{opRHS} annihilating the right-hand side of~\eqref{eq.5} has been computed
by exploiting $\partial$-finite closure properties in the spirit of
\thm{thm.clprop}.  We continue with \thm{thm.sumint} which establishes the
closure of holonomic functions with respect to sums and integrals; for its
proof, we once again refer to~\cite{Zeilberger90,Koutschan09}.
\begin{theorem}\label{thm.sumint}
  Let the function $f$ be holonomic with respect to $D_{\!x}$ (resp. $S_n$). Then also
  $\int_a^b f\rmd x$ (resp. $\sum_{n=a}^b f$) is holonomic.
\end{theorem}

\begin{example}
We continue the discussion from Example~\ref{ex.Gegen1} by again considering
the Gegenbauer polynomials
\begin{equation}\label{eq.Gegen}
  C_m^{(\nu)}(x) := \sum_{k=0}^m F[x,m,\nu,k]
\end{equation}
where
\begin{mma}
  \In |F|[x\blank ,m\blank ,\nu\blank ,k\blank ]:=\frac{|P|[2\nu,m]}{m!}
    \frac{|P|[-m,k]\,|P|[m+2\nu,k]}{|P|[\nu+1/2,k]\,k!)}\Bigl(\frac{1-x}{2}\Bigr)^k\\
\end{mma}
\noindent
with
\begin{mma}
  \In |P|[x\blank ,k\blank ]:=|Pochhammer|[x,k]\\
\end{mma}
\noindent
This time we want to derive the annihilator of $C_m^{(\nu)}(x)$ from its
definition~\eqref{eq.Gegen}. For this purpose, we compute annihilating operators of
the hypergeometric term $F[x,m,\nu,k]$ in telescoping form:
\begin{mma}
  \In |CreativeTelescoping|[|F|[x,m,\nu,k],|S|[k]-1,\{|Der|[x],|S|[m]\}] // |Factor|\\
  \Out \Bigl\{\{-(x-1) (x+1) D_{\!x}+(m+1) S_{m}+x (-(m+2 \nu)),\linebreak
    \phantom{\Bigl\{\{}(m+2) S_{m}^2-2 x (m+\nu+1) S_{m}+(m+2 \nu)\},\linebreak
    \phantom{\Bigl\{}\Bigl\{-\frac{k (2 k+2 \nu-1)}{k-m-1},\frac{2 k (2 k+2 \nu-1) (m+\nu+1)}{(k-m-2) (k-m-1)}\Bigr\}\Bigr\}\\
\end{mma}
\noindent
The output has to be interpreted as follows:
\begin{multline}\label{eq.T1}
  \bigl(-(-1+x) (1+x) D_x + (m+1)S_m - (m+2\nu)x\bigr) F[x,m,\nu,k] =\\
  (S_k-1) \frac{k(2k+2\nu-1)}{k-m-1} F[x,m,\nu,k] \qquad
\end{multline}
and
\begin{multline}\label{eq.T2}
  \bigl((m+2)S_m^2 - 2(m+\nu+1)xS_m + (m+2\nu)\bigr) F[x,m,\nu,k] =\\
  -(S_k-1) \frac{2k(m+\nu+1)(2k+2\nu-1)}{(k-m-2)(k-m-1)} F[x,m,\nu,k].
\end{multline}
Note that the relations~\eqref{eq.T1} and~\eqref{eq.T2} can be easily verified
(even without using a computer). To compute them, the package
\cmd{HolonomicFunctions} employs non-commutative Gr\"obner bases;
the monomial order is deduced from the order in which the operators
are given. Indeed, by changing in the input the order of
\cmd{Der}[$x$] and \cmd{S}[$m$], one obtains a different result:
\begin{mma}
  \In |CreativeTelescoping|[|F|[x,m,\nu,k],|S|[k]-1,\{|S|[m],|Der|[x]\}] // |Factor|\\
  \Out \Bigl\{\{(m+1) S_{m}-(x-1) (x+1) D_{\!x}+x (-(m+2\nu)),\linebreak
    \phantom{\Bigl\{\{}-(x-1) (x+1) D_{\!x}^2-(2\nu+1) x D_{\!x}+m (m+2 \nu)\},\linebreak
    \phantom{\Bigl\{}\Bigl\{-\frac{k (2 k+2 \nu-1)}{k-m-1},-\frac{k (2 k+2 \nu-1)}{x-1}\Bigr\}\Bigr\}\\
\end{mma}
\noindent
This repeats~\eqref{eq.T1}, but computes another purely differential relation
\begin{multline}\label{eq.T3}
  \bigl(-(-1+x)(1+x)D_x^2 - (2\nu+1)xD_x + m(m+2\nu)\bigr) F[x,m,\nu,k] = \\
  (S_k-1) \frac{k(2k+2\nu-1)}{x-1} F[x,m,\nu,k].
\end{multline}
Summing \eqref{eq.T1}, \eqref{eq.T2}, and \eqref{eq.T3} with respect to $k$
from $0$ to $\infty$ gives the well-known shift/differential relations for the
Gegenbauer polynomials.
\end{example}

These computations were done in the operator algebra $\OO=\F[S_m,S_k,D_x]$ with
$\F=\Q(\nu,m,k,x)$. Let us include in addition the shift operator~$S_\nu$:
\begin{mma}
  \In |CreativeTelescoping|[|F|[x,m,\nu,k],|S|[k]-1,\{|S|[m],|S|[\nu],|Der|[x]\}]\\
  \Out \Bigl\{\{2 \nu S_\nu-x D_{\!x}+(-m-2 \nu),(m+1) S_m+(1-x^2) D_{\!x}+(-m x-2 \nu x),\linebreak
    \phantom{\Bigl\{\{}(1-x^2) D_{\!x}^2+(-2 \nu x-x) D_{\!x}+(m^2+2 m \nu)\},\linebreak
    \phantom{\Bigl\{}\Bigl\{-\frac{k}{x-1},\frac{-2 k^2-2 k \nu+k}{k-m-1},\frac{-2 k^2-2 k \nu+k}{x-1}\Bigl\}\Bigl\}\\
\end{mma}
\noindent
We see that summing the resulting telescoping relations with respect to $k$
from $0$ to~$\infty$, gives the generators of the annihilating ideal
\cmd{annG} computed in \mmaref{Out}{\ref{annG}}.

Finally we note that for the $q$-case we need to consider the Gegenbauer
polynomials in the (equivalent) form:
\[
  C_m^{(\nu)}\bigl(\cos(\theta)\bigr) := \sum_{k=0}^m G[x,m,\nu,k]
\]
where
\begin{mma}
  \In G[x\blank ,m\blank ,\nu\blank ,k\blank ]:=\frac{|P|[\nu,k]\,|P|[\nu,m-k]}{k!(m-k)!}A^{m-2k}\\
\end{mma}
\noindent
with $x=\cos(\theta)$, $A=e^{i\theta}$ and $P[\nu,k]$ being defined as the
Pochhammer symbol $(\nu)_k$, as in \mmaref{In}{\ref{srec}}.
In the $q$-context we will be interested to compute annihilating operators
containig shifts in~$m$ and~$\nu$, and, as above, in telescoping form with
respect to $S_k-1$. More precisely, in the next section we will compute a
$q$-version of
\begin{mma}
  \In |CreativeTelescoping|[G[x,m,\nu,k],|S|[k]-1,\{|S|[m],|S|[\nu]\}] // |Factor|\\
  \Out \Bigl\{\{A \bigl(A^2+1\bigr) (m+1) S_{m}-(A-1)^2 (A+1)^2 \nu S_{\nu}-2 A^2 (m+2 \nu), \linebreak
    \phantom{\Bigl\{\{}-(A-1)^2 (A+1)^2 \nu (\nu+1) S_{\nu}^2+\nu \bigl(A^4 m+A^4 \nu+A^4-2 A^2 m-6 A^2 \nu-{} \linebreak
    \phantom{\Bigl\{\{}4 A^2+m+\nu+1\bigr) S_{\nu}+A^2 (m+2 \nu) (m+2 \nu+1)\},\linebreak
    \phantom{\Bigl\{}\Bigl\{\frac{A^2 k (k-m-\nu) \bigl(A^2 k-A^2 m+A^2 \nu-A^2-k+m+\nu+1\bigr)}{\nu (k-m-1)},\linebreak
    \phantom{\Bigl\{\Bigl\{}\Bigl(A^2 k (-k+m+\nu) \bigl(A^2 k^2-A^2 k m+A^2 k \nu-A^2 m \nu-k^2+k m+k \nu+{} \linebreak
    \phantom{\Bigl\{\Bigl\{\Bigl(}2k-m-\nu-1\bigr)\Bigr)\mathrel{\Big{\slash}}\Bigl(\nu (\nu+1)\Bigr)\Bigr\}\Bigr\}\\
\end{mma}

\section{The Ismail-Zhang Formula}
\label{sec.IZ}

An important classical expansion formula is the expansion of the plane wave in
terms of ultraspherical polynomials $C_m^{(\nu)}(x)$, also called Gegenbauer
polynomials:
\[
  e^{irx} = \left(\frac2r\right)^\nu \Gamma(\nu) \sum_{m=0}^\infty i^m (\nu+m)\, J_{\nu+m}(r)\, C_m^{(\nu)}(x).
\]
Ismail and Zhang~\cite[(3.32)]{IsmailZhang94} had found the following $q$-analog of this formula:
\begin{multline}\label{eq.IZ}
  \Eq(x;i\omega) = \frac{(q;q)_\infty \,\omega^{-\nu}}{(q^\nu;q)_\infty\, (-q\omega^2;q^2)_\infty}\\
  \times\sum_{m=0}^\infty i^m (1-q^{\nu+m}) \, q^{m^2/4} J_{\nu+m}^{(2)}(2\omega;q)\, C_m(x;q^\nu|q),\qquad
\end{multline}
where $J_{\nu+m}^{(2)}(2\omega;q)$ is Jackson's $q$-Bessel function defined by
\[
  J_\nu^{(2)}(z;q) = \frac{(q^{\nu+1};q)_\infty}{(q;q)_\infty} \sum_{n=0}^\infty q^{(\nu+n)n} \frac{(-1)^n (z/2)^{\nu+2n}}{(q;q)_n (q^{\nu+1};q)_n}.
\]
In the Ismail-Zhang formula~\eqref{eq.IZ}, Jackson's second $q$-analog of the
Bessel function $J_\nu(z)$ appears; the remaining ingredients, the basic
exponential function $\Eq(x;i\omega)$ and the $q$-Gegenbauer polynomials
$C_m(x;q^\nu|q)$, are explained subsequently. There are several proofs of the
Ismail-Zhang formula; see the books~\cite{Ismail05} and~\cite{SuslovBook} for
references and for the embedding of the formula in a broader context.  In this
section we present a new, computer-assisted proof of~\eqref{eq.IZ}.

\subsection{The basic exponential function}

The basic exponential function $\Eq(x;i\omega)$, as well as its more general
version $\Eq(x,y;i\omega)$, was introduced by Ismail and Zhang
\cite{IsmailZhang94} and satisfies numerous important and also beautiful
properties.  For illustrative reasons we choose to introduce $\Eq(x;i\omega)$
via the basic cosine and sine functions: For $x=\cos(\theta)$ and $|\omega|<1$
we define:
\[
  \Eq(x;i\omega) := C_q(x;\omega) + i\,S_q(x;\omega)
\]
where the basic cosine function $C_q(x;\omega)$ is defined as
\[
  C_q(x;\omega) := \frac{(-\omega^2;q^2)_\infty}{(-q\omega^2;q^2)_\infty}
    \sum_{j=0}^\infty \frac{(-qe^{2i\theta};q^2)_j\,(-qe^{-2i\theta};q^2)_j}{(q;q^2)_j\,(q^2;q^2)_j} (-\omega^2)^j,
\]
and the basic sine function $S_q(x;\omega)$ as
\[
  S_q(x;\omega) := \frac{(-\omega^2;q^2)_\infty}{(-q\omega^2;q^2)_\infty} \frac{2q^{1/4}\omega}{1-q} \cos(\theta)
    \sum_{j=0}^\infty \frac{(-qe^{2i\theta};q^2)_j\,(-qe^{-2i\theta};q^2)_j}{(q^3;q^2)_j\,(q^2;q^2)_j} (-\omega^2)^j.
\]
It is not difficult to check that
\begin{align*}
  \lim_{q\to1-} C_q\bigl(x;\omega(1-q)/2\bigr) &= \cos(\omega x) \\
  \lim_{q\to1-} S_q\bigl(x;\omega(1-q)/2\bigr) &= \sin(\omega x)  
\end{align*}
In the following we shall use the abbreviation $A=e^{i\theta}$, as before,
and the following short-hand notation for the \cmd{qPochhammer} command:
\begin{mma}
  \In |qP|=|qPochhammer|;\\
\end{mma}
\noindent
Consequently, the input \cmd{qP}[$-w^2,q^2$] stands for
$(-\omega^2;q^2)_\infty$, and \cmd{qP}[$-qA^2,q^2,j$] for
$(-qe^{2i\theta};q^2)_j$.  The continuous $q$-ultraspherical ($q$-Gegenbauer)
polynomials $C_m(x;q^\nu|q),\,x=\cos(\theta)$, are defined as
\[
  C_m(\cos\theta;\beta|q) := \sum_{k=0}^m \frac{(\beta;q)_k\,(\beta;q)_{m-k}}{(q;q)_k\,(q;q)_{m-k}} e^{i(m-2k)\theta}.
\]

To prove~\eqref{eq.IZ} we compute annihilating operators representing
$q$-difference equations for the left- and right-hand sides,
respectively. First we derive a $q$-shift equation for the $q$-Cosine.  This
is done analogously to the treatment of the right-hand side of~\eqref{eq.6}:
\begin{mma}
  \In |CreativeTelescoping|\Bigl[\frac{|qP|[-\omega^2,q^2]}{|qP|[-q\omega^2,q^2]}
    \frac{|qP|[-qA^2,q^2,j]\,|qP|[-q/A^2,q^2,j]}{|qP|[q,q^2,j]\,|qP|[q^2,q^2,j]}(-\omega^2)^j,\hspace{-21pt}\linebreak
    |QS|[J,q^j]-1,|QS|[\omega,q^w]\Bigr] // |Factor|\\
  \Out \Bigl\{\bigl\{A^2 \bigl(q^2 \omega ^2+1\bigr) S_{\omega,q}^2+(A^4 q^2 \omega ^2-A^2 q-A^2+q^2 \omega ^2) S_{\omega,q}+
    A^2 q \bigl(q \omega ^2+1\bigr)\bigr\},\linebreak
    \phantom{\Bigl\{}\Bigl\{\frac{A^2 (J-1) (J+1) \bigl(J^2-q\bigr) \bigl(q \omega ^2+1\bigr)}{\omega ^2+1}\Bigr\}\Bigr\}\\
\end{mma}
\noindent
Denoting by $c_j(\omega)$ the summand in the $q$-cosine series and in view of
$S_{\omega,q}^jf(\omega)=f(q^j\omega)$, this output means that
\begin{multline*}
  A^2 \bigl(q^2 \omega ^2+1\bigr)c_j(q^2 \omega) + (A^4 q^2 \omega ^2-A^2 q-A^2+q^2 \omega ^2)c_j(q \omega) +
  A^2 q \bigl(q \omega ^2+1\bigr)c_j(\omega) = \\
  - \Delta_j \frac{A^2 (J-1) (J+1) \bigl(J^2-q\bigr) \bigl(q \omega ^2+1\bigr)}{\omega ^2+1} c_j(\omega),
\end{multline*}
where $J=q^j$. Summing the right-hand side from $j=0$ to $j=\infty$ gives
\[
  -\frac{q A^2 (1+q \omega^2)}{1+\omega^2} c_\infty(\omega) +
  \frac{A^2 (-1+q^0) (1+q^0) (q^0-q) (1+q\omega^2)}{1+\omega^2} c_0(\omega) = 0.
\]
Hence
\begin{mma}
  \In |annCos|=\%[[1]]\\
  \Out \bigl\{A^2 \bigl(q^2 \omega ^2+1\bigr) S_{\omega,q}^2+(A^4 q^2 \omega ^2-A^2 q-A^2+q^2 \omega ^2) S_{\omega,q}+
    A^2 q \bigl(q \omega ^2+1\bigr)\bigr\}\\
\end{mma}
\noindent
annihilates $C_q(x;\omega)$.  An annihilator, resp. $q$-difference equation,
for the $q$-Sine is derived analogously:
\begin{mma}
  \In |Factor|\Bigl[|CreativeTelescoping|\Bigl[\mmi\frac{|qP|[-\omega^2,q^2]\,2q^{1/4}\omega}{|qP|[-q\omega^2,q^2](1-q)}|Cos|[\theta]\linebreak
    \!\frac{|qP|[-q^2A^2,q^2,j]\,|qP|[-q^2/A^2,q^2,j]}{|qP|[q^3,q^2,j]\,|qP|[q^2,q^2,j]} (-\omega^2)^j,|QS|[J,q^j]-1,|QS|[\omega,q^w]\Bigr]\Bigr]\\
  \Out \Bigl\{\bigl\{A^2 \bigl(q^2 \omega ^2+1\bigr) S_{\omega,q}^2+(A^4 q^2 \omega ^2-A^2 q-A^2+q^2 \omega ^2) S_{\omega,q}+
    A^2 q \bigl(q \omega ^2+1\bigr)\bigr\},\linebreak
    \phantom{\Bigl\{}\Bigl\{\frac{A^2 (J-1) (J+1) q \bigl(J^2 q-1\bigr) \bigl(q \omega ^2+1\bigr)}{\omega ^2+1}\Bigr\}\Bigr\}\\
  \In |annSin|=\%[[1]]\\
  \Out \bigl\{A^2 \bigl(q^2 \omega ^2+1\bigr) S_{\omega,q}^2+\bigl(A^4 q^2 \omega ^2-A^2 q-A^2+q^2 \omega ^2\bigr) S_{\omega,q}+
    A^2 q \bigl(q \omega ^2+1\bigr)\bigr\}\\
\end{mma}
Finally we exploit the $q$-holonomic closure properties; more precisely, in
view of Theorem~\ref{thm.clprop}(i) we ``add'' the $q$-difference equations
for the $q$-cosine function and $i$ times the $q$-sine function to obtain a
$q$-difference equation for $\Eq(x;i\omega)$. The latter is the generator of
the annihilating ideal of $\Eq(x;i\omega)$:
\begin{mma}
  \In |annLHS|=|DFinitePlus|[|annCos|,|annSin|]\\
  \Out \bigl\{\bigl(A^2 q^2 \omega ^2+A^2\bigr) S_{\omega,q}^2+\bigl(A^4 q^2 \omega ^2-A^2 q-A^2+q^2 \omega ^2\bigr) S_{\omega,q}+
    \bigl(A^2 q^2 \omega ^2+A^2 q\bigr)\bigr\}\\
\end{mma}
\noindent
The result is not surprising: since $C_q(x;\omega)$ and $S_q(x;\omega)$
satisfy the same $q$-difference equation (compare \cmd{annSin} with
\cmd{annCos}), also their linear combination satisfies the same
equation. Conversely, the operator above annihilates any linear combination
\[
  c_1 C_q(x;\omega) + c_2 S_q(x;\omega)
\]
where $c_1$ and $c_2$ are constants, i.e., independent of $\omega$. The order
of \cmd{annLHS} is~$2$, hence one derives explicit expressions for the $c_i$
by picking the coefficients of $\omega^0$ and $\omega^1$ in $\Eq(x;i\omega)$,
respectively:
\begin{equation}\label{eq.7}
  c_1 = 1,\quad
  c_2 = \frac{2q^{1/4}}{1-q}\cos(\theta).
\end{equation}
Summarizing, $\Eq(x;i\omega)$ is uniquely determined by the $q$-difference
operator \cmd{annLHS} and the initial values~\eqref{eq.7}.

\subsection{An annihilator for the Ismail-Zhang series}

To compute an annihilating operator for the right-hand side of~\eqref{eq.IZ},
we algorithmically exploit $\pa$-finite, resp. $q$-holonomic, closure
properties as described in Section~\ref{sec.ann}.  Let us first compute
generators of the ideal of operators annihilating
$C_m(\cos\theta;q^\nu|q)$. Recall $x=\cos(\theta)$ and $A=e^{i\theta}$. In
addition, we will use the abbreviations
\[
  K = q^k,\quad M = q^m,\quad N = q^n,\quad V = q^\nu,\text{ and}\quad \omega = q^w.
\]
\begin{mma}
  \In |annqGegenbauer|=\linebreak
    |CreativeTelescoping|\Bigl[\frac{|qP|[q^\nu,q,k]\,|qP|[q^\nu,q,m-k]}{|qP|[q,q,k]\,|qP|[q,q,m-k]}A^{m-2k},\linebreak
    |QS|[|K|,q^k]-1,\{|QS|[M,q^m],|QS|[V,q^\nu],|QS|[\omega,q^w]\}\Bigr][[1]] // |Factor|\\
  \Out \bigl\{S_{\omega,q}-1,-A \bigl(A^2+1\bigr) V (M q-1) S_{\!M,q}+(V-1) \bigl(A^2-V\bigr) \bigl(A^2 V-1\bigr) S_{V,q}+{}\linebreak
    \phantom{\bigl\{}A^2 (V+1) \bigl(M V^2-1\bigr),(V-1) (q V-1) \bigl(A^2-q V\bigr) \bigl(A^2 q V-1\bigr) S_{V,q}^2-{}\linebreak
    \phantom{\bigl\{}(V-1) \bigl(A^4 M q^2 V^2-A^4 q V-A^2 M q^3 V^3-A^2 M q^2 V^3+A^2 q+A^2+{}\linebreak
    \phantom{\bigl\{}M q^2 V^2-q V\bigr) S_{V,q}-A^2 q \bigl(M V^2-1\bigr) \bigl(M q V^2-1\bigr)\bigr\}\\
\end{mma}
The algorithmic method to compute \cmd{annqGegenbauer} follows the
creative telescoping strategy described in Section~\ref{sec.ann}. In
contrast to the $q=1$ case, here we also include the shift with respect
to~$\omega$ which in the output gives rise to an additional, trivial generator
$S_{\omega,q}-1$. This is done in order to be able to execute all required
closure property computations in one common operator algebra. For further
details see~\cite{Koutschan09,Koutschan10b}; there is also an on-line
description of the \cmd{CreativeTelescoping} procedure in the
\cmd{HolonomicFunctions} package:
\begin{mma}
 \In ?|CreativeTelescoping|\\
\end{mma}
\vskip 2pt\noindent\hskip 24pt\fbox{\parbox{\mmaboxwidth}{\small
CreativeTelescoping[f, delta, \{op1, \dots, opk\}] or
CreativeTelescoping[ann, delta, \{op1, \dots, opk\}] computes creative
telescoping relations for the given function f (resp. the given
$\pa$-finite ideal ann annihilating some function~f). In
particular it returns \{\{q1, \dots, qm\}, \{r1, \dots, rm\}\}, two lists of
OrePolynomials such that qj + delta*rj is in the annihilator of f for
all $1\leq j\leq m$. The polynomials qj form a Groebner basis in the
rational Ore algebra with generators op1, \dots, opk whereas the rj's
live in the Ore algebra with generators Join[OreOperators[delta],
\{op1, \dots, opk\}] (resp. the Ore algebra of ann). For summation
(w.r.t.~n) set delta to S[n]--1 or Delta[n], and in the q-case to
QS[qn,q$\>\pow\>$n]--1; for integration (w.r.t.~x) set delta to Der[x].}}
\vskip 8pt

In an analogous fashion we compute generators of the annihilating ideal of
the $q$-Bessel function $J_\nu^{(2)}(2\omega;q)$:
\begin{mma}
  \In |annqBesselJ|=\linebreak
    |CreativeTelescoping|\Bigl[q^{(\nu+n)n}\frac{|qP|[q^{\nu+1},q]\,(-1)^n\omega^{\nu+2n}}{|qP|[q,q]\,|qP|[q,q,n]\,|qP|[q^{\nu+1},q,n]},\linebreak
    |QS|[|N|,q^n]-1,\{|QS|[M,q^m],|QS|[V,q^\nu],|QS|[\omega,q^w]\}\Bigr][[1]]\\
  \Out \bigl\{(-V \omega -\omega) S_{V,q}+(q \omega ^4+q \omega ^2+\omega ^2+1) S_{\omega,q}+(\omega ^2-V),S_{\!M,q}-1,\linebreak
    (q^5 V \omega ^4+q^3 V \omega ^2+q^2 V \omega ^2+V) S_{\omega,q}^2+(q^2 V \omega ^2+q V \omega ^2-V^2-1) S_{\omega,q}+V\bigr\}\\
\end{mma}
\noindent
The annihilating ideal of $h_1(\omega,m,n):=i^m(1-q^{\nu+m})$ is obtained as follows:
\begin{mma}
  \In |annh1|=|Annihilator|[\mmi^m(1-q^{\nu+m}),\{|QS|[M,q^m],|QS|[V,q^\nu],|QS|[\omega,q^w]\}]\\
  \Out \bigl\{S_{\omega,q}-1,(M V-1) S_{V,q}+(1-M q V),(M V-1) S_{\!M,q}+(i-i M q V)\bigr\}\\
  \label{annh1}
\end{mma}
\noindent
Note that in fact it is trivial to compute the generators of this ideal, just
consider the quotients
\[
  \frac{h_1(q\omega,m,n)}{h_1(\omega,m,n)},
  \quad \frac{h_1(\omega,m+1,n)}{h_1(\omega,m,n)},
  \quad\text{and}\quad \frac{h_1(\omega,m,n+1)}{h_1(\omega,m,n)},
\]
which, after simplification, yield rational functions, whose numerators and
denominators appear as coefficients in the first-order operators of
\mmaref{Out}{\ref{annh1}}. The reason for this is that
$h_1(\omega,m,n)$ is actually a $q$-hypergeometric term. When trying to
compute the annihilating ideal of $h_2(m):=q^{m^2/4}$ by the \cmd{Annihilator}
command, the package is trapped by the factor $\frac14$ in the exponent and
delivers the fourth-order operator $S_{\!M,q}^4-q^4M^2$. Although this is
correct, in the sense that it is a left multiple of the minimal-order
annihilating operator, it is not the operator we wish to work with. Instead,
we write down the annihilator of $h_2(m)$ by hand, and convert its generators
to Ore polynomials that live in the same Ore algebra as \cmd{annh1}:
\begin{mma}
  \In |annh2|=|ToOrePolynomial|[\{|QS|[V,q^\nu]-1,|QS|[\omega,q^w]-1,\linebreak
    |QS|[M,q^m]^2-qM\},|OreAlgebra|[|QS|[M,q^m],|QS|[V,q^\nu],|QS|[\omega,q^w]]]\\
  \Out \bigl\{S_{V,q}-1,S_{\omega,q}-1,S_{\!M,q}^2-M q\bigr\}\\
\end{mma}
\noindent
The annihilating ideal of $h_1(\omega,m,n)h_2(m)=i^m(1-q^{\nu+m})q^{m^2/4}$, is
obtained by applying the closure property ``multiplication'', see
Theorem~\ref{thm.clprop}(ii):
\begin{mma}
  \In |annh1h2|=|DFiniteTimes|[|annh1|,|annh2|]\\
  \Out \bigl\{S_{\omega,q}-1,(M V-1) S_{V,q}+(1-M q V),(M V-1) S_{\!M,q}^2+(M^2 q^3 V-M q)\bigr\}\\
\end{mma}
\noindent
We continue by applying the same closure property again, in order to obtain an
annihilating ideal of
$i^m(1-q^{\nu+m})q^{m^2/4}J_{\nu+m}^{(2)}(2\omega;q)C_m(\cos\theta;q^\nu|q)$; here
we use the previously computed annihilators \cmd{annqBesselJ} and
\cmd{annqGegenbauer} of the $q$-Bessel function and the $q$-Gegenbauer
polynomials, respectively, together with discrete substitution as described in
Theorem~\ref{thm.clprop}(vi):
\begin{mma}
  \In |annSmnd|=|DFiniteTimes|[|annh1h2|,\linebreak
    |DFiniteSubstitute|[|annqBesselJ|,\{\nu\to \nu+m\}],|annqGegenbauer|];\\
\end{mma}
The output list consists of three annihilating operators, and it would require
about two pages to display them. The output \cmd{annSmnd} is of the form:
\begin{multline}
 \bigl\{(1+q^2\omega^2+q^3\omega^2+q^5\omega^4)MVS_{\omega,q}^2+(qMV\omega^2+q^2MV\omega^2-M^2V^2-1)S_{\omega,q} + MV,\\
 (-A^2 M V \omega^2+\ldots-q^5 A^2 M^5 V^9 \omega^2)S_{V,q}^2+(A^2 M V \omega+\ldots+q^6 A^2 M^5 V^8 \omega^5)S_{V,q} S_{\omega,q} + {}\!\!\\
 (-q A^2 M^2 V^2 \omega-\ldots+q^5 A^2 M^5 V^8 \omega^3)S_{V,q}+(-A^2+\ldots+q^7 A^2 M^6 V^8 \omega^4)S_{\omega,q} + {}\\
 (A^2 M V-\ldots+q^6 A^2 M^6 V^8 \omega^2),\\
 (-A^2 V^2 \omega^2+\ldots+q^4 A^2 M^5 V^5 \omega^2)S_{M,q}^2+(q A^2 M V \omega-\ldots-q^4 A^2 M^4 V^7 \omega^5)S_{V,q} S_{\omega,q} + {}\!\!\!\\
 (-q^2 A^2 M^2 V^2 \omega+\ldots-q^3 A^2 M^4 V^7 \omega^3)S_{V,q}+(-A^2+\ldots-q^6 A^2 M^5 V^7 \omega^4)S_{\omega,q} + {}\\
 (A^2 M V-\ldots-q^5 A^2 M^5 V^7 \omega^2)\bigr\}
\end{multline}
The next step in the process of constructing an annihilating ideal of the
right-hand side of~\eqref{eq.IZ} consists in ``doing the sum''
\[
  \sum_{m=0}^\infty i^m(1-q^{\nu+m})\,q^{m^2/4}J_{\nu+m}^{(2)}(2\omega;q)\,C_m(\cos\theta;q^\nu|q).
\]
However, applying the \cmd{CreativeTelescoping} command, as we did before,
does not deliver any result within a reasonable amount of time. By inspecting
the leading monomials of \cmd{annSmnd}---they are $S_{\omega,q}^2$,
$S_{V,q}^2$, and $S_{M,q}^2$---we find that the holonomic rank of
\cmd{annSmnd} is~$8$, which is relatively large and which explains the failure
of the first attempt. Luckily there exists an algorithm~\cite{Koutschan10c}
that is more efficient in such situations, but whose drawback is that
sometimes it is not able to deduce the correct denominator of the output. The
current summation problem is such an example, and therefore we give the
correct denominator with an additional option (it can be found by looking at
the leading coefficients of \cmd{annSmnd} plus some trial and error). The
computation then takes about 20 seconds and for better readability we suppress
parts of the output.
\begin{mma}
  \In |annSumRHS|=|FindCreativeTelescoping|[|annSmnd|,|QS|[M,q^m]-1,\linebreak
    |Denominator|\to (M^2 V^2-1)(q^2 M^2 V^2-1)(q^4 M^2 V^2-1)]\\
  \Out \Bigl\{\bigl\{(V-1) S_{V,q}+\omega,(-q^5 A^2 \omega ^4-q^3 A^2 \omega ^2-q^2 A^2 \omega ^2-A^2) S_{\omega,q}^2 + {} \linebreak
    \phantom{\Bigl\{\bigl\{}(-q^2 A^4 V \omega^2+q A^2 V+A^2 V-q^2 V \omega ^2) S_{\omega,q}-q A^2 V^2\bigl\},\Bigl\{ \cdots \Bigr\}\Bigr\}\\
  \label{annSumRHS}
\end{mma}
\noindent
Finally we obtain the annihilating ideal of the right-hand side of~\eqref{eq.IZ}:
\begin{mma}
  \In |annRHS|=|DFiniteTimes|[|annSumRHS|[[1]],|Annihilator|[\linebreak
      |qP|[q,q]/(\omega^\nu|qP|[q^\nu,q]\,|qP|[(-q)\omega^2,q^2]),\{|QS|[V,q^\nu],|QS|[\omega,q^w]\}]]\\
  \Out \bigl\{S_{V,q}-1,(q^2 A^2 \omega ^2+A^2) S_{\omega,q}^2+(q^2 A^4 \omega ^2-qA^2-A^2+q^2 \omega ^2) S_{\omega,q} +
    (q^2 A^2 \omega ^2+qA^2)\bigr\}\\
\end{mma}
\noindent
Comparison with the left-hand side of~\eqref{eq.IZ}:
\begin{mma}
  \In |annLHS|\\
  \Out \bigl\{(q^2 A^2 \omega ^2+A^2) S_{\omega,q}^2+(q^2 A^4 \omega ^2-qA^2-A^2+q^2 \omega ^2) S_{\omega,q} +
    (q^2 A^2 \omega ^2+qA^2)\bigr\}\\
  \label{annLHS}
\end{mma}
To complete the proof we have to incorporate initial conditions. To this end
we convert the $q$-shift equation to an equivalent version which is in the
format of a $q$-differential equation.  This is supported by the command
\cmd{QSE2DE} from the \cmd{qGeneratingFunctions} package; in order to invoke
it, we need to convert the operator \cmd{annLHS} in \mmaref{Out}{\ref{annLHS}}
to a standard $q$-shift equation:
\begin{mma}
  \In |qSeq|=|ApplyOreOperator|[|annLHS|,f[\omega]]/.q\pow (a\blank .\,m+b\blank .)\to q^b\omega^a\\
  \Out \bigl\{ (q^2 A^2 \omega ^2 + q A^2) f[\omega] + (q^2 A^4 \omega ^2 - q A^2 - A^2 + q^2 \omega ^2) f[q\omega] +
    (q^2 A^2 \omega ^2 + A^2) f[q^2\omega] \bigr\}\\
  \In |QSE2DE|[|qSeq|,f[\omega]]\\
  \Out \bigl\{q (A^2+1)^2 f[\omega] + q (q-1) (A^4 + q A^2 + A^2 + 1) \omega f'[\omega] +
   (q-1)^2 (q^2 \omega ^2+1) A^2 f''[\omega]= 0\bigr\}\\
  \label{qde}
\end{mma}
In this equivalent form, $f'(\omega):=D_qf(\omega)$ refers to the $q$-derivative defined on (formal) power series as
\[
  D_{\!q}\sum_{n=0}^\infty a_n\omega^n := \sum_{n=1}^\infty a_n\frac{q^n-1}{q-1}\omega^{n-1}.
\]
Now our proof can be completed as follows: let $l(\omega)$ and $r(\omega)$
denote the left and right sides of~\eqref{eq.IZ}, respectively. Above we have
shown that both $l(\omega)$ and $r(\omega)$ satisfy the $q$-differential
equation \mmaref{Out}{\ref{qde}}.  So what is left to show is that
\[
  l(0) = r(0) = 1
  \quad\text{and}\quad
  l'(0) = r'(0) = \frac{2iq^{1/4}\cos(\theta)}{1-q}.
\]
But this, in view of the definition of $D_q$, amounts to comparing the
coefficients of $\omega^0$ and $\omega^1$, respectively, in the Taylor
expansions of $l(\omega)$ and $r(\omega)$. Owing to the definitions of the
functions involved and noticing that $\omega^{-\nu}$ is cancelling out, this
task is an easy verification.

\section{Conclusion}
\label{sec.conclude}

As expressed in the Introduction, a major objective of this article is to
popularize the holonomic systems approach in the field of $q$-series and basic
hypergeometric functions. In the case studies we presented, RISC software
written in Mathematica was used.  With respect to $q$-summation one can find
various packages written in Maple or in other computer algebra systems; with
regard to the more general $q$-holonomic setting (operator algebras,
non-commutative Gr\"obner basis methods, etc.) we point explicitly to the Maple
package \cmd{Mgfun} by F.~Chyzak~\cite{Chyzak98}.

We want to conclude with a few remarks on the fact that computing an
annihilating operator for the series side of the Ismail-Zhang
formula~\eqref{eq.IZ} is leading to the frontiers of what is computationally
feasible today. As already pointed out, ongoing research is trying to push
frontiers further by the design of new constructive methods
like~\cite{Koutschan10c}.  Formulas like~\eqref{eq.IZ} or the very well-poised
basic hypergeometric series ${}_{10}W_9$ in~\cite{IsmailRainsStanton15}, which
has been successfully treated by the \cmd{HolonomicFunctions} package, provide
excellent challenges and inspirations for such algorithmic developments.

\bibliographystyle{plain}

\end{document}